\documentclass[letterpaper,fleqn]{cas-sc}

\usepackage[numbers]{natbib}
\usepackage{listings}
\usepackage{hyperref}
\usepackage{pdflscape}
\newcommand{\rev}[1]{\textcolor{black}{#1}}
\newcommand{\revthree}[1]{\textcolor{black}{#1}}
\def\tsc#1{\csdef{#1}{\textsc{\lowercase{#1}}\xspace}}
\tsc{WGM}
\tsc{QE}
\usepackage{lineno}

\begin{document}
\let\WriteBookmarks\relax
\def\floatpagepagefraction{1}
\def\textpagefraction{.001}

\newcommand{\loveland}[1]{\textcolor{black}{#1}}

\shorttitle{Efficacy of reduced order source terms for a coupled wave-circulation model in the Gulf of Mexico}    

\shortauthors{Loveland \emph{et al.}}  

\title [mode = title]{Efficacy of reduced order source terms for a coupled wave-circulation model in the Gulf of Mexico}  



%

\author[1]{Mark Loveland}[orcid=0000-0002-2164-2884]

\cormark[1]


\ead{markloveland@utexas.edu}

\ead[url]{https://www.researchgate.net/profile/Mark-Loveland-2}

\credit{Methodology, Software, Validation, Writing - Original Draft}

\affiliation[1]{organization={Oden Institute for Computational Engineering and Sciences, University of Texas at Austin},
            addressline={Oden Institute for Computational Engineering and Sciences, The University of Texas at Austin. 201 E. 24th St.}, 
            city={Austin},
            postcode={78712}, 
            state={Texas},
            country={USA}}

\author[2]{Jessica Meixner}
\credit{Methodology, Writing - Review \& Editing}
\affiliation[2]{organization={NWS/NCEP/Environmental Modeling Center, National Oceanic and Atmospheric Administration (NOAA)},
            addressline={}, 
            city={College Park},
            postcode={}, 
            state={MD},
            country={USA}}
 
\author[1,3,4]{Eirik Valseth}[orcid=0000-0001-6940-4191]



\credit{Methodology, Writing - Review \& Editing}

\affiliation[3]{organization={The Department of Data Science, The Norwegian University of Life Science},
            addressline={Drøbakveien 31}, 
            city={Ås},
            postcode={1433}, 
            country={Norway}}
\affiliation[4]{organization={Department of Scientific Computing and Numerical Analysis, Simula Research Laboratory},
            addressline={Kristian Augusts gate 23}, 
            city={Oslo},
            postcode={0164}, 
            country={Norway}}

\author[1]{Clint Dawson}[orcid=0000-0001-7273-0684]
\credit{Resources, Supervision, Project administration}
\cortext[1]{Corresponding author}



\begin{abstract}
During hurricanes, coupled wave-circulation models are critical tools for public safety. The standard approach is to use a high fidelity circulation model coupled with a wave model \rev{that} uses the most advanced source terms. As a result, the models can be computationally expensive and so this study investigates the potential consequences of using simplified (reduced order) source terms within the wave model component of the coupled wave-circulation model. The trade-off between run time and accuracy with respect to observations is quantified for a set of two storms that impacted the Gulf of Mexico, Hurricane Ike and Hurricane Ida. Water surface elevations as well as wave statistics (significant wave height, peak period, and mean wave direction) are compared to observations. The usage of the reduced order source terms yielded significant savings in computational cost. Additionally, relatively low amounts of additional error with respect to observations during the simulations with reduced order source terms \rev{are observed in our computational experiments}. However, large changes in global model outputs of the wave statistics were observed based on the choice of source terms particularly near the track of each hurricane.
\end{abstract}


\begin{highlights}
\item ADCIRC+SWAN models  Hurricane Ida and with various source term choices for SWAN
\item Reduced order source terms (1st, 2nd gen) reduce run time by nearly half
\item Source terms had larger impact on wave statistics than water surface elevations
\end{highlights}


\maketitle








\section{Introduction}\label{sec:intro}

During tropical cyclones, the destructive impacts of wind waves and storm surge can be hazardous for local populations, infrastructure, flora, and fauna. Accurate numerical models are important tools in preparation for such events so that forecasts can be made and emergency measures may be taken to avoid unnecessary damage. It has been shown that the interaction between wind waves and the surge plays a significant role in the overall storm surge levels via wave-induced set up and wave-induced set down~\cite{longuet-higgins_stewart_1962}. This observed interaction has resulted in the development of coupled spectral wave and circulation models, which model both wind waves and storm surge respectively, to improve model accuracy. In this study we will focus on coupled models that use a spectral wave model governed by the Wave Action Balance Equation (WAE), and a circulation model governed by the Shallow Water Equations. There are many operational-scale models of this kind that include coupling between circulation and wind waves to predict storm surge and wave statistics including ADCIRC+SWAN~\cite{DIETRICH201145}, ADCIRC+STWAVE~\cite{ADCIRCSTWAVE,ADCIRCSTWAVE2}, AdH+STWAVE~\cite{mcalpin2020new}, ADCIRC+WAVEWATCHIII~\cite{ADCIRCWW3}, and SCHISM\rev{+}WWMIII~\cite{roland2012fully}. Spectral wave models are expensive and in particular the computation of the source terms within the models can take a large portion of the computing load. \loveland{The inclusion of the spectral wave model increases total run times by as much as triple that of a standalone circulation model.}~\cite{Rolandthesis,Meixnerthesis,komen_cavaleri_donelan_hasselmann_hasselmann_janssen_1994,state_of_art,MONBALIU2003133,Janssen2008,Khandekar_1989,roland2012fully}.

The source terms within the context of the WAE are both complex and  empirical due to the complex nature of the physics \rev{the WAE} is trying to approximate and is still an active area of research~\cite{komen_cavaleri_donelan_hasselmann_hasselmann_janssen_1994,state_of_art}. Since the latter half of the 20th century, the general format of the source terms for the WAE has remained relatively unchanged, and is often written as a sum of wind input ($S_{\rev{\text{in}}}$), nonlinear interaction ($S_{\rev{\text{nl}}}$), and dissipation($S_{\rev{\text{diss}}}$), though the specific parameterizations of these are always evolving~\cite{Khandekar_1989}. \loveland{The choice of source terms can  influence not only the accuracy of results but can also  influence computational cost~\cite{komen_cavaleri_donelan_hasselmann_hasselmann_janssen_1994,state_of_art,SWAMP_1985}. In this study we seek to quantify the trade-off between the accuracy and the computational cost due to source term choices, specifically in the coupled wave-circulation setting.}

\loveland{Operational wave-circulation models often utilize the latest, most accurate source terms available for the given wave model~\cite{DIETRICH201145,ADCIRCSTWAVE,mcalpin2020new,ADCIRCWW3,roland2012fully}. These source term packages are referred to as 3rd generation \rev{(Gen3)} source terms because they explicitly approximate the full nonlinear interactions contained in the aforementioned $S_{\rev{\text{nl}}}$ term~\cite{SWAMP_1985,komen_cavaleri_donelan_hasselmann_hasselmann_janssen_1994}. The two main physical processes that $S_{\rev{\text{nl}}}$ accounts for are called three-wave interactions (triads) and four-wave interactions (quadruplets)~\cite{komen_cavaleri_donelan_hasselmann_hasselmann_janssen_1994,Janssen2008,holthuijsen_2007}. These processes are both known to be computationally expensive to approximate accurately and therefore even the most cutting edge \rev{Gen3} source term packages rely on very crude approximations~\cite{state_of_art}. 
The most common methods to approximate the triads and quadruplets are the lumped triad approximation (LTA)~\cite{eldeberky1997nonlinear} and discrete interaction approximation (DIA)~\cite{Hasselmann_1985} respectively. Even these rough approximations can still be computationally expensive, and so part of this study is to see how much cost can be saved and what accuracy is sacrificed if these processes are either neglected or simplified within the coupled wave-circulation setting.}

\loveland{It is also noted that there is still an ongoing debate on the physical accuracy namely of the  source term packages relying on the DIA~\cite{InconsistentSpectralEvolutioninOperationalWaveModelsduetoInaccurateSpecificationofNonlinearInteractions}. This has motivated work on finding parametric relations for waves generated by hurricanes to reduce model cost while still incorporating effects from waves~\cite{YUROVSKAYA2023102184}. Furthermore, there has been investigation into using reduced order or parametric models for the spectral waves to force circulation models~\cite{BOYD2021107192}. This work seeks to look further into this direction by looking into the usage of 1st and 2nd generation \rev{(Gen1 and Gen2)} source terms for the wave model within the coupled wave-circulation setting. \rev{Gen1 and Gen2} source terms are characterized by either negligence or severe simplification of the nonlinear source term $S_{\rev{\text{nl}}}$~\cite{SWAMP_1985}. For a more detailed exposition into the specifics of the distinction between \rev{Gen1, Gen2, and Gen3} source terms we refer to~\cite{SWAMP_1985,komen_cavaleri_donelan_hasselmann_hasselmann_janssen_1994}.
}

\loveland{In this study, ADCIRC+SWAN will be used to investigate the trade-off between computational performance and accuracy of \rev{Gen1, Gen2, and Gen3} source term parameterizations within SWAN. In particular, this work will focus on the impacts of utilizing the reduced-order physics of the \rev{Gen1 and Gen2}  source term choices within ADCIRC+SWAN when compared to the more commonly used \rev{Gen3} source terms. The accuracy of the models will be quantified by looking at both SWAN wave statistic output compared to real buoys as well as ADCIRC water surface elevation (WSE) output compared to real gauges. The computational performance of the models will be quantified by tabulating wall clock times.} 

Following this introductory section, in Section~\ref{sec:Methods}, ADCIRC+SWAN will briefly be described along with the source term packages used within SWAN. Next, in Section~\ref{sec:scenarios} the specific storms used in this study, Hurricanes Ike and  Ida, will be discussed and presented in detail. Then in Section~\ref{sec:details}, the specific configuration for each ADCIRC+SWAN \rev{simulation} will be outlined in detail. The results of all of the simulations are shown and analyzed against field data in Section~\ref{sec:Results}. Finally, conclusions and comments on potential future directions of work are drawn in Section~\ref{sec:con}.

\section{Model Details}\label{sec:Methods}

\subsection{The ADCIRC+SWAN Model}\label{subsec:ADCIRC+SWAN}
The shallow water model that is coupled to SWAN is called the Advanced Circulation Model (ADCIRC)~\cite{luettich2004formulation}. ADCIRC is a finite element model that solves the shallow water equations (SWE) on unstructured triangular meshes and is used often to predict tides and storm surges~\cite{westerink2008basin}. Both standalone ADCIRC and the coupled model ADCIRC+SWAN have been validated in many realistic scenarios such as \revthree{Hurricane Katrina, Hurricane Rita, and Hurricane Ike, and are} currently used operationally for forecasting and hindcasting ~\cite{DIETRICH201145,dietrich2012performance,fleming2008real,hope2013hindcast}.

The coupled model (ADCIRC+SWAN) works through a tight, sequential coupling set up where ADCIRC is run for some time and the resulting WSE and mean water velocities are passed as inputs to SWAN. SWAN then runs given this information and then outputs wave radiation stress, which is added as forcing into the momentum equations of ADCIRC~\cite{DIETRICH201145}. Both ADCIRC and SWAN are run on the same unstructured mesh, which eliminates the need for interpolation of WSE, water velocities, wave radiation stresses, bathymetry, or wind inputs.

ADCIRC obtains the WSE and depth-averaged velocities by solving the Generalized Wave Continuity Equation (GWCE) version of the SWE, which is obtained by differentiation of the continuity equation of the SWE in time and addition of the momentum equations of the SWE~\cite{luettich2004formulation}. ADCIRC subsequently solves the weak form of the GWCE using the standard Bubnov-Galerkin method with linear polynomial basis functions. \rev{The details of the weak form are written in several published papers~\cite{luettich2004formulation,dietrich2012performance}. When the impact of wind waves are considered in ADCIRC, a term in the right hand side of the momentum equations is added that approximates the applied force due to wind waves. The added terms are often referred to as wave stresses and can be written as:}

\begin{equation}\label{eqn:wavestress}
    \begin{split}
        \tau_{sx,\rev{\text{waves}}} = -\frac{\partial S_{xx}}{\partial x} - \frac{\partial S_{xy}}{\partial y}, \\
        \tau_{sy,\rev{\text{waves}}} = -\frac{\partial S_{xy}}{\partial x} - \frac{\partial S_{yy}}{\partial y},
    \end{split}
\end{equation}
where $S_{xx}, S_{xy}$, and $S_{yy}$ are the wave radiation stresses computed with SWAN output and defined as:
\begin{equation}
    \begin{split}
        S_{xx} = \rho g \int_{-\pi}^{\pi} \int_{\sigma_{\rev{\text{min}}}}^{\sigma_{\rev{\text{max}}}} (n \textrm{cos}^2(\theta) + n -\frac{1}{2}) \sigma N d\sigma d\theta, \\
        S_{xy} = \rho g \int_{-\pi}^{\pi} \int_{\sigma_{\rev{\text{min}}}}^{\sigma_{\rev{\text{max}} }}n\textrm{sin}(\theta) \textrm{cos}(\theta) \sigma N d\sigma d\theta, \\
        S_{yy} = \rho g \int_{-\pi}^{\pi} \int_{\sigma_{\rev{\text{min}}}}^{\sigma_{\rev{\text{max}}}} (n \textrm{sin}^2(\theta) + n -\frac{1}{2}) \sigma N d\sigma d\theta, 
    \end{split}
\end{equation}
where \rev{$\rho$ is the density of water, $g$ is the gravitational constant, }$N$ is the action balance, $\sigma, \theta$ represents relative radial frequency and direction of the wave spectrum respectively. The variable $n$ is related to propagation velocity and is defined as:
\begin{equation}\label{eqn:n}
    n = \frac{1}{2}\left(1 + \frac{2kd}{\sinh{(2kd)}}\right),
\end{equation}
where $k$ is the wavenumber magnitude defined through the dispersion relation~\cite{holthuijsen_2007}\rev{, and $d$ is total water depth \revthree{that} comes from ADCIRC by adding the WSE to the bathymetric value}.

The SWAN model is defined through the WAE. The WAE is a linear, scalar-valued, time-dependent, hyperbolic equation in 4 dimensions with a varying (both in 4-D space and time), non-divergence free velocity field,  ($\mathbf{c}(x,y,\sigma,\theta,t)$), which can be determined independently of the \rev{solution variable,} $N(x,y,\sigma,\theta,t)$. This is different than most conservation laws such as Navier-Stokes and related transport equations because typically the propagation velocity is just a function of the unknown i.e. $\mathbf{c}(N)$. \rev{The specific definition of the WAE and its derivation can be found in several publications and textbooks~\cite{holthuijsen_2007,SWAMP_1985,AThirdGenerationModelforWindWavesonSlowlyVaryingUnsteadyandInhomogeneousDepthsandCurrents,Janssen2008,komen_cavaleri_donelan_hasselmann_hasselmann_janssen_1994,leblond1981waves,TheWAMModel}. The SWAN model uses a finite difference scheme to solve the WAE for action balance on unstructured grids~\cite{Booij_1999,holthuijsen1988wave,swantech}. It is important to note how the ADCIRC and SWAN models depend on one another. Through~\eqref{eqn:wavestress}, the solution of ADCIRC depends on the solution of SWAN via wave stresses. The SWAN model solution uses the ADCIRC WSE through~\eqref{eqn:n} and ADCIRC's depth-averaged velocities in order to compute the propagation velocity, $\mathbf{c}(x,y,\sigma,\theta,t)$~\cite{holthuijsen_2007,swantech}. The SWAN model may also depend on the WSE and depth-averaged velocities in order to compute contributions to the source term in the WAE, $S(N)$~\cite{swantech}}.

\subsection{SWAN Source Term Packages}\label{subsec:SWAN_source}

In the definition of the \rev{WAE,} 
the source term $S(N)$ \rev{in its right hand side is}   arbitrary. In practice, the source term $S$ can take many forms but in general it can be thought of as a sum of three key sources/sinks:
\begin{equation}\label{eqn:Hass}
    S = S_{\rev{\text{in}}} + S_{\rev{\text{diss}}} + S_{\rev{\text{nl}}},
\end{equation}
where $S_{\rev{\text{in}}}$ represents any contribution to the spectrum due to wind input and typically takes the form as a sum of linear and exponential wave growth:
\begin{equation}\label{eqn:Sin}
    S_{\rev{\text{in}}} = \alpha + \beta E(x,y,\sigma,\theta,t),
\end{equation}
where $\alpha,\beta$ are tunable parameters. $S_{\rev{\text{diss}}}$ represents any change in the energy spectrum due to dissipation that can include things like whitecapping and surf zone breaking. The most commonly used dissipation source terms are whitecapping, bottom friction, and depth-induced breaking:
\begin{equation}
    S_{\rev{\text{diss}}} = S_{\rev{\text{wc}}} + S_{\rev{\text{bf}}} + S_{\rev{\text{br}}}.
\end{equation}
$S_{\rev{\text{nl}}}$ represents changes in the spectrum due to nonlinear wave interactions.  In the following subsections we summarize the forms of the source term in the 3rd generation ST6 package as well as 1st and 2nd generation source terms that are implemented in SWAN and will be used in this study. A more general discussion of some of the available source terms in SWAN can be found in the SWAN technical manual~\cite{swantech}.
\loveland{ The ST6 package was chosen to represent the 3rd generation source terms because this is the latest available in the ADCIRC+SWAN software, and the specific implementations of the 1st and 2nd generation source term packages were chosen because they were the only ones readily available within the SWAN source code.}

\subsubsection{Full Source Terms (ST6)}
In the last decade, beginning with the work of Ardhuin, Babanin and many others~\cite{Ardhuin2010}, the latest third generation source term package has changed to become semi-empirical in an aim to agree better with observed data and specifically to be more robust to situations with extreme weather.  This work culminated into new source terms for $S_{\rev{\text{in}}}$ and $S_{\rev{\text{wc}}}$, and is often referred to as ST6, which was first fully defined in a paper by Rogers \emph{et al.} in 2012~\cite{Rogers2012}. This source term package is available and widely used in both SWAN and WAVEWATCH III and will be used in this study. \rev{The specific version of SWAN used in this study is the latest available within the ADCIRC+SWAN software package and is version v41.31. The ST6 source terms in this version of SWAN are identical to those descrived in the Rogers paper~\cite{Rogers2012}. \revthree{It is noted that for the nonlinear source terms, $S_{nl}$, the DIA is used for the four-wave interactions and three-wave interactions are neglected by default. In this study, three-wave interactions will be omitted.}}

\subsubsection{Reduced Order Source Terms}
In addition to the ST6 source term package, which represents a 3rd generation type, SWAN contains a \rev{Gen1} and \rev{Gen2} source package as formulated in a study from Holthuijsen et. al.~\cite{holthuijsen1988wave}. These source packages in SWAN are defined in the technical manual~\rev{\cite{swantech}. These source term configurations will neglect any explicit approximations for nonlinear interactions, $S_{nl}$, and the form that the other terms, $S_{in}$ and $S_{diss}$, take are significantly simplified.} 
Now that we have established how exactly ADCIRC and SWAN influence one another as well as the source term configurations that will be used, we will describe the specific test cases used in this study.

\section{Scenario Descriptions}\label{sec:scenarios}
\subsection{Hurricane Ike}
Hurricane Ike is selected as a test case because there is plenty of data available for comparisons and because this specific scenario has already been validated with ADCIRC+SWAN in~\cite{hope2013hindcast}. Hurricane Ike made landfall on September 13th, 2008 near the City of Galveston, Texas. Hurricane Ike was a category 4 hurricane with maximum sustained winds of 110 mph at the time of landfall in Galveston~\cite{uscommerce_2017}. Hurricane Ike produced billions of dollars in damages and induced deadly storm surge across the upper Texas and southwest Louisiana coasts that claimed dozens of lives~\cite{zane2011tracking}. 

Hurricane Ike in particular was chosen as a test case for this study because high levels of storm surge were generated and the fact that wave interaction in some areas resulted in up to 50\% changes in modeled storm surge~\cite{hope2013hindcast}. Thus, we know a-priori that the results from the SWAN model will have significant impacts on the accuracy of the resulting ADCIRC water levels and water velocities. \loveland{More specifically, wave-induced set up from Ike is known to have been particularly large in the vicinity of the hurricane track and so we can expect a high sensitivity of the coupled wave-circulation output to the choices of source terms in the wave model.}

\subsection{Hurricane Ida}
Hurricane Ida was a category 4 storm that made landfall on August 29, 2021 near Port Fourchon, Louisiana~\cite{hanchey2021notes}. The storm had an intense wind field with sustained winds of over 150 mph causing billions of dollars in damage and several lives to be lost.

A main reason that Hurricane Ida was chosen as a second test case is because the storm produced large amounts of surge, recorded as high as 4.3 m in some areas~\cite{Idanotes}. Additionally, the hurricane made landfall in a separate part of the Gulf of Mexico than Hurricane Ike, thus improving the robustness of our study to more than just a single case study. Additionally, because this storm is more recent there is ample data available to compare simulation results to. \loveland{The data that will be used to evaluate errors in the model include NOAA gauge data for WSE and NOAA buoy data for wave statistics.}

\section{Implementation Details}\label{sec:details}

\subsection{Grids}
The computational grid used in the Hurricane Ike scenario comes from the work of \revthree{Kennedy} et. al. and  consists of 6,675,517 elements and 3,352,598 nodes with resolution ranging from the order of km in the deep ocean up to 20 m or less along the Texas coast where Ike made landfall~\cite{Kennedy2011}. \loveland{ Bathymetric data that defined the mesh was taken from a combination of satellite data sets and digital elevation models near the coast as specified in~\cite{Kennedy2011,hope2013hindcast}.}

\begin{figure}[h!]
    \centering
    \includegraphics[width=0.55\textwidth]{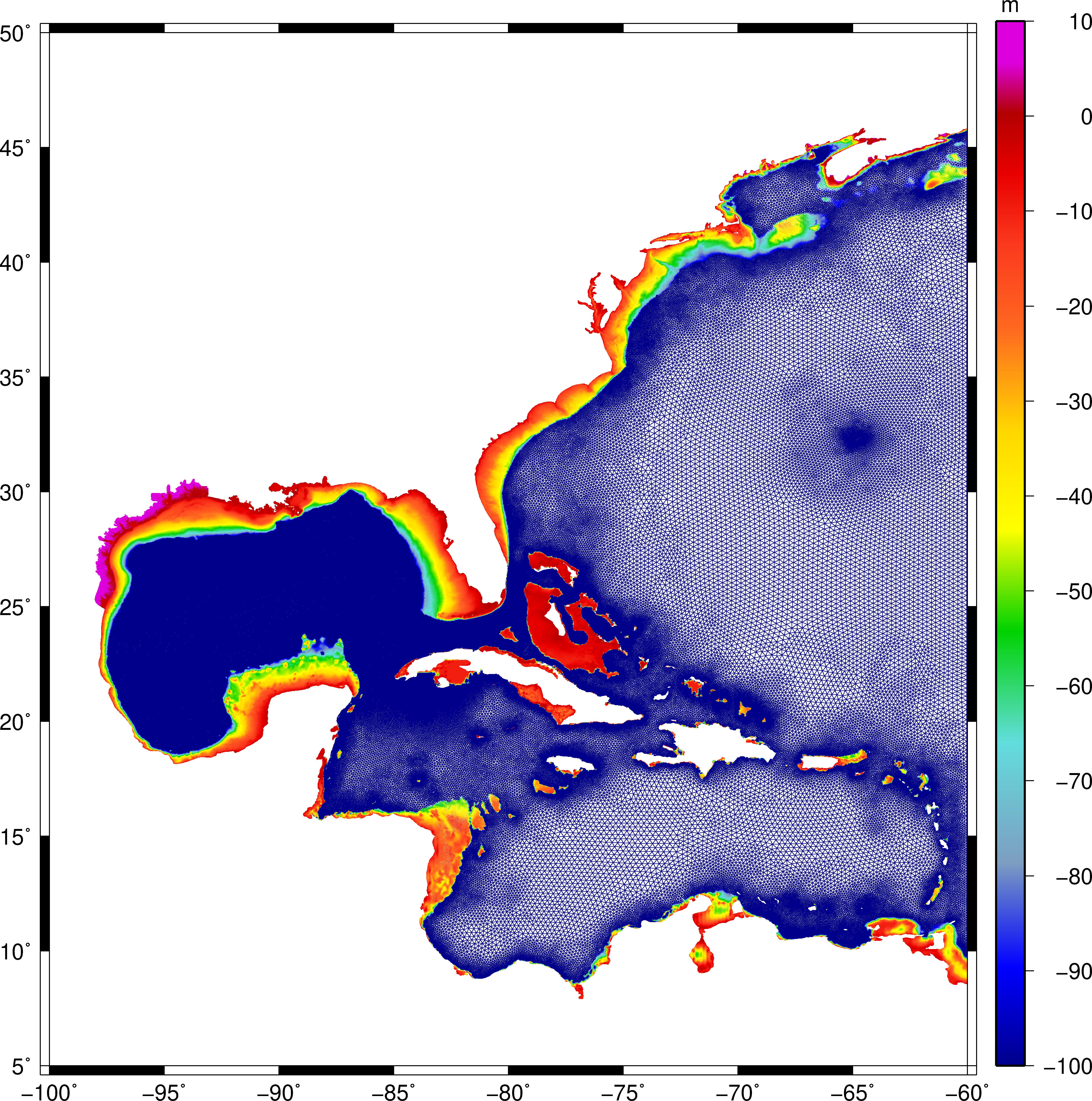}
    \caption{\rev{Computational mesh for Hurricane Ike, bathymetry cut off at 100m.} \revthree{The mesh for Hurricane Ida has similar extents except with higher levels of refinement near the Louisiana coastline as opposed to the Texas coastline.}}
    \label{fig:ikemesh}
\end{figure}

For the Hurricane Ida scenario, a different mesh is used with high refinement specifically along the Louisiana coast \revthree{that comes from the work of Dietrich et. al. in 2008 \cite{dietrich2011hurricane}}. The mesh consists of 3,102,441 elements and 1,593,485 nodes with resolution in a similar range to that of the mesh used in Hurricane Ike. In this case, resolution around the Louisiana coast is as fine as 20 m while the remainder of the Gulf Coast has refinement of anywhere from the order of 100 m and deep ocean is as sparser on the order of km.
\loveland{The mesh for Ida is constructed from similar data sources to the Ike mesh. The mesh used for Ida is different than Ike because each mesh has particularly high resolution in the area where the storms made landfall, which is critical in achieving the accuracy of any circulation-wave model.} \rev{Both meshes consequently have the same extent covering the ocean west of the $60^{\circ}$ meridian from Venezuela to Canada, see \textbf{Figure~\ref{fig:ikemesh}}).  }

In both scenarios, the spectral domain for SWAN is 36 cells in direction, $\theta$, and 40 cells in frequency, $\sigma$. The directions have uniform spacing of $10^{\circ}$ and range across the full circle, $-180^{\circ}$ to $180^{\circ}$ and the frequency goes from  $0.031384 Hz - 2.55Hz$  with logarithmic spacing such that $f_{i+1} = \gamma f_i$ where $\gamma=1.1$ is a constant. \loveland{These specifications for the spectral grid may be slightly higher than operational forecast models but were chosen to allow for sufficient refinement of the SWAN model because the focus of this study is the impact of source term choices on the accuracy of the coupled wave-circulation model.}

\subsection{Forcing and Boundary Conditions}\label{subsec:forcing}
The input winds and pressure for the Hurrricane Ike scenario are validated meteorology hindcasts from Ocean\rev{weather} Inc. (OWI). \revthree{Because the OWI wind fields are not publicly available for Hurricane Ida, the best track winds from the NHC's HURDAT2~\rev{\cite{HURDAT2}} dataset are used to force pressure and winds~\rev{\cite{idaHurdat}}}. This is also a similar wind format used during forecasting and thus provides an opportunity to assess a wind product used in such applications. \loveland{HURDAT2 gives parametric information about the location and size of the hurricane and the Generalized Asymmetric Holland Model \rev{(GAHM)~\cite{gao2018surface}} is used to approximate the actual wind and pressure fields that force ADCIRC+SWAN. } \rev{We refer to the ADCIRC Wiki page\footnote{\url{https://wiki.adcirc.org/Generalized\_Asymmetric\_Holland\_Model\#cite\_note-1}} for further details on the GAHM.}

For both scenarios, conditions for the open ocean boundary at the $60^{\circ}$ meridian for ADCIRC are generated from the TPXO global tide model~\cite{EfficientInverseModelingofBarotropicOceanTides}. To avoid inducing artificial oscillations into the simulation a 30 day tidal spin up of ADCIRC without winds and hence, no waves, is supplied as initial conditions to the ADCIRC+SWAN model in both scenarios. For Hurricane Ike, total simulation time is 8 days and 18 hours beginning on September 5, 2008 at 12:00 GMT and ending September 14, 2008 at 06:00 GMT. For Hurricane Ida the total run time is 9 days and 6 hours beginning at August 26, 2021 at 12:00 GMT and ending at September 4, 2021 at 18:00 GMT.
For the SWAN wave model, the initial conditions are set to 0 and the open boundary conditions for SWAN are 0.  

This study is conducted by running simulations of both Hurricane Ike and Hurricane Ida, each with identical forcing inputs while changing only the source term implementations in SWAN. ADCIRC is run without SWAN as a control and then 1st Generation (Gen1), 2nd Generation (Gen2), and the latest ST6 (Gen3) packages, as described in Section~\ref{sec:details}, are run. The coupling interval between ADCIRC and SWAN happens every 600 seconds for both scenarios. \revthree{The required input file that specifies the SWAN configuration within ADCIRC+SWAN, the fort.26 file, can be found for each source term configuration in Appendix~\ref{ap:Source}. Furthermore, the full set of inputs, outputs, and run instructions for all of the simulations are publicly available on DesignSafe~\cite{DesignSafePaper}, and can be accessed through the DOI present in the following citation~\cite{data_designsafe}. }

\subsection{Post Processing}
\begin{figure}[h!]
    \centering
    \includegraphics[width=\textwidth]{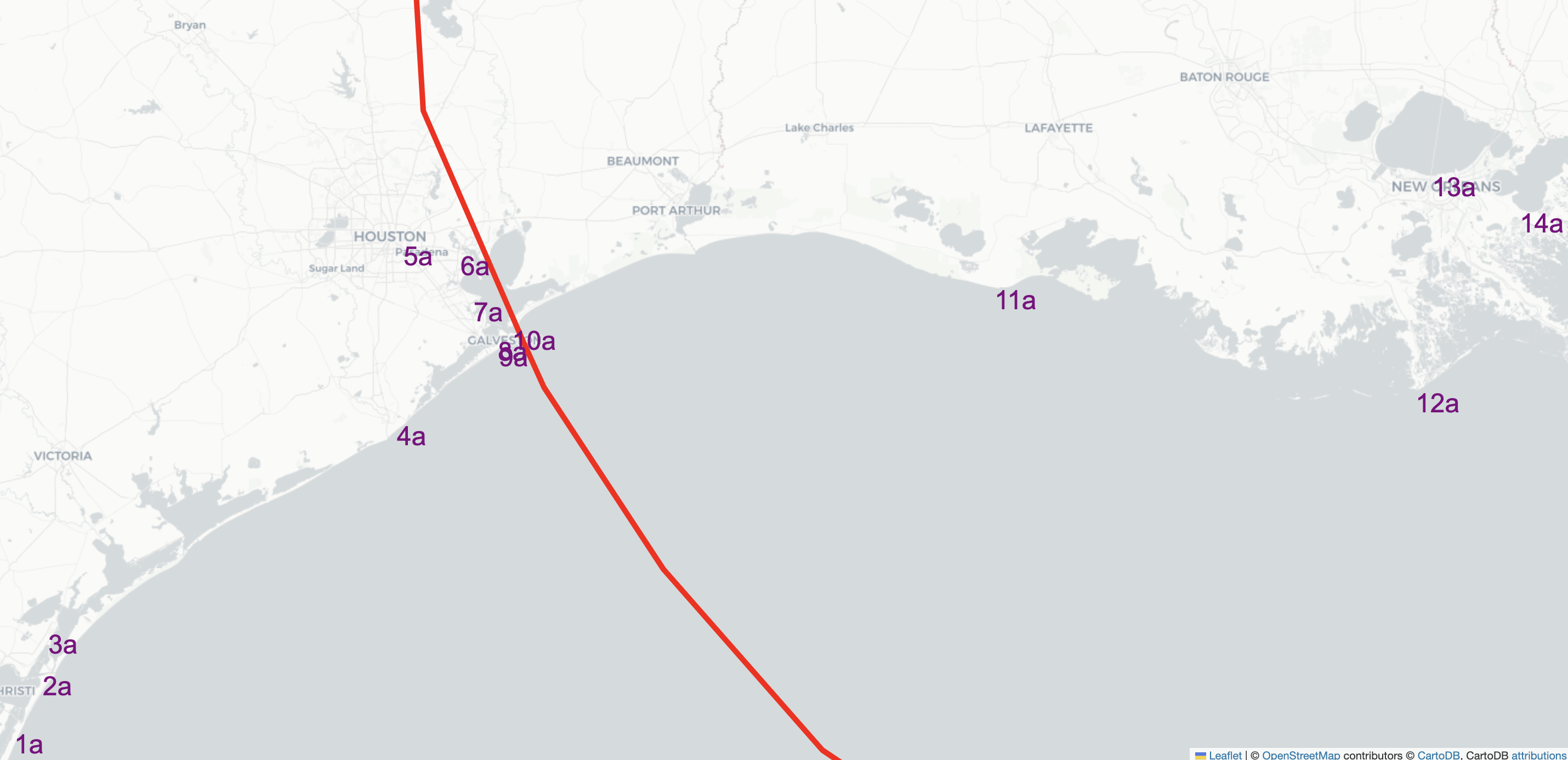}
    \caption{All water surface elevation gauges used during the Hurricane Ike test case. Hurricane Ike track in red.}
    \label{fig:ele_gauge_all}
\end{figure}

\begin{figure}[h!]
    \centering
    \includegraphics[width=\textwidth]{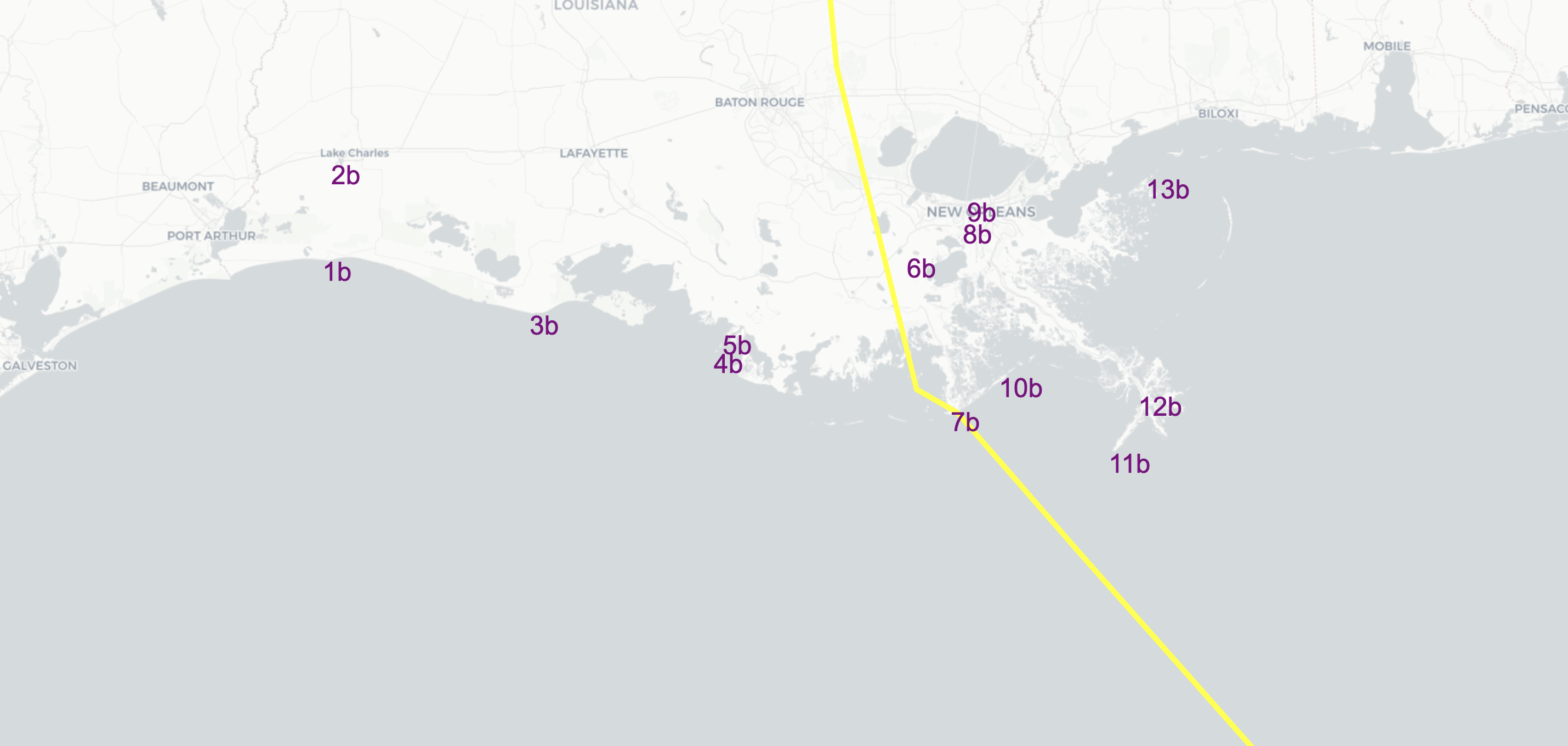}
    \caption{Locations of NOAA water surface elevation gauges used to compare for Hurricane Ida. Hurricane Ida track in yellow.}
    \label{fig:Ida_gauge_locs}
\end{figure}

\begin{figure}[h!]
    \centering
    \includegraphics[width=\textwidth]{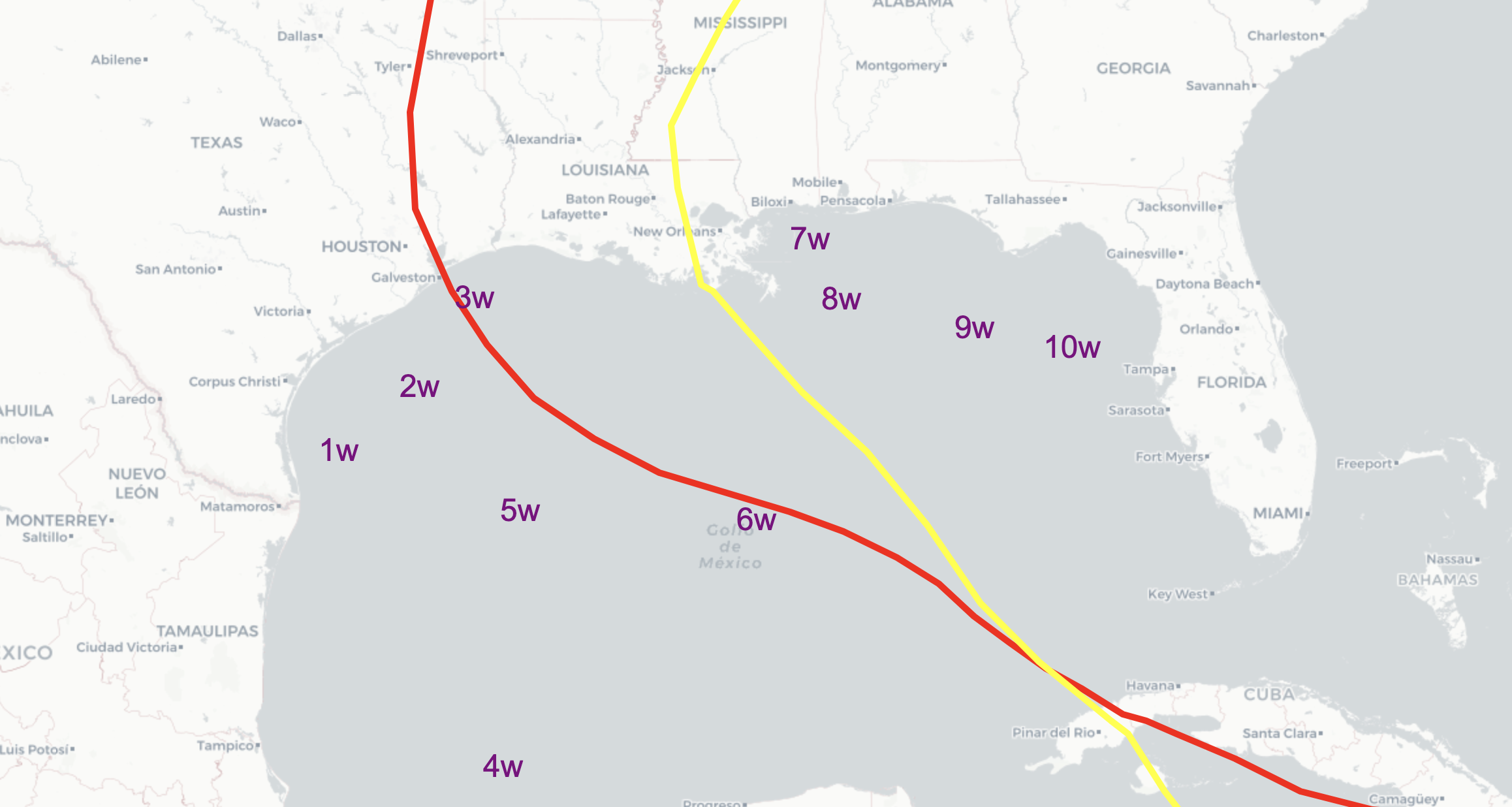}
    \caption{Wave buoy locations that are used in both of the scenarios. Hurricane Ike track in red and Hurricane Ida track in yellow.}
    \label{fig:wave_gauges}
\end{figure}

The ensuing ADCIRC+SWAN water levels are compared to available measurements from NOAA gauges and significant wave height, mean wave period, and mean wave direction are compared to available NOAA buoys. A summary of the gauge locations for Hurricane Ike can be found in \textbf{Table~\ref{tab:Elevation_Stations}} and can be seen on the maps in ~\textbf{Figure~\ref{fig:ele_gauge_all}}. The gauge locations for Hurricane Ida can be found in~\textbf{Table~\ref{tab:Ida_Elevation_Stations}} and the locations can be seen on the map in~\textbf{Figure~\ref{fig:Ida_gauge_locs}}. A summary of the buoy locations used in both scenarios can be found in \textbf{Table~\ref{tab:Wave_Stations}} and their locations are shown on a map in \textbf{Figure~\ref{fig:wave_gauges}}. \revthree{All gauges and buoys are numbered from West to East.} 
\begin{table}[h!]
\centering
\caption{\label{tab:Elevation_Stations}  Water surface elevation stations used to compare ADCIRC+SWAN results for Hurricane Ike.}
\begin{tabular}{|l|l|l|l|l|}
\hline
{Gauge no.} & { Gauge Name} &{NOAA ID} & {Latitude} & {Longitude}   \\
\hline
1a & Bob Hall Pier Corpus Christi & 8775870 & 27.5800 & -97.2167 \\
\hline
2a & Port Aransas & 8775237 & 27.8404 & -97.0730\\
\hline
3a & Rockport & 8774770 & 28.0217 & -97.0467\\
\hline
4a & USCG Freeport & 8772447 & 28.9428 & -95.3025\\
\hline
5a & Manchester Houston & 8770777 & 29.7247 & -95.2656\\
\hline
6a & Morgans Point & 8770613 & 29.6817 & -94.9850\\
\hline
7a & Eagle Point & 8771013 & 29.4813 & -94.9172 \\
\hline
8a & Galveston Pier 21 & 8771450 & 29.3100 & -94.7933\\
\hline
9a & Galveston Pleasure Pier & 8771510 & 29.2849 & -94.7894\\
\hline
10a & Galveston Bay Entrance North Jetty & 8771341 & 29.3576 & -94.7260\\
\hline
11a & Freshwater Canal Locks & 8766072 & 29.5341 & -92.3082 \\
\hline
12a & Port Fourchon & 8762075 & 29.0848 & -90.1985\\
\hline
13a & New Canal Station & 8761927 & 30.0272 & -90.1134\\
\hline
14a & Shell Beach & 8761305 & 29.8681 & -89.6733\\
\hline
\end{tabular}
\end{table}
%
%
 %
 %
\begin{table}[h!]
\centering
\caption{\label{tab:Ida_Elevation_Stations}  Water surface elevation stations used to compare ADCIRC+SWAN results for Hurricane Ida.}
\begin{tabular}{|l|l|l|l|l|}
\hline
{Gauge no.} & { Gauge Name} & {NOAA ID} & {Latitude} & {Longitude}   \\
\hline
1b & Calcasieu Pass & 8768094 & 29.7683 & -93.3433\\
\hline
2b & Bulk Terminal & 8767961 & 30.1900 & -93.3000\\
\hline
3b & Freshwater Canal Locks & 8766072 & 29.5341 & -92.3082\\
\hline
4b & Eugene Island & 8764314 & 29.3667 & -91.3833\\
\hline
5b & LAWMA Amerada Pass & 8764227 & 29.4500 & -91.3383\\
\hline
6b & West Bank 1 & 8762482 & 29.7838 & -90.4200\\
\hline
7b & Port Fourchon & 8762075 & 29.1142 & -90.1993 \\
\hline
8b & Carrollton & 8761955 & 29.9333 & -90.1350 \\
\hline
9b & New Canal Station & 8761927 & 30.0272 & -90.1133\\
\hline
10b & Grand Isle & 8761724 & 29.2633 & -89.9567 \\
\hline
11b & Pilots Station East & 8760922 & 28.9316 & -89.4067\\
\hline
12b & Pilottown, LA & 8760721 & 29.1793 & -89.2588 \\
\hline
13b & Shell Beach & 8761305 & 30.1267 & -89.2217 \\
\hline
\end{tabular}
\end{table}
\begin{table}[h!]
\centering
\caption{\label{tab:Wave_Stations}  Wave buoy info.}
\begin{tabular}{|l|l|l|l|}
\hline
{Buoy no.} & { Buoy Name} & {Latitude} & {Longitude}   \\
\hline
1w & NBDC 42020 & 26.9680 & -96.6930 \\
\hline
2w & NBDC 42019 & 27.9100 & -95.3450 \\
\hline
3w & NBDC 42035 & 29.2320 & -94.4130 \\
\hline
4w & NBDC 42055 & 22.1240 & -93.9410 \\
\hline
5w & NBDC 42002 & 26.0550 & -93.6460 \\
\hline
6w & NBDC 42001 & 25.9190 & -89.6740 \\
\hline
7w & NBDC 42007 & 30.0900 & -88.7690 \\
\hline
8w & NBDC 42040 & 29.2070 & -88.2370 \\
\hline
9w & NBDC 42039 & 28.7870 & -86.0070 \\
\hline
10w & NBDC 42036 & 28.5010 & -84.5080 \\
\hline
\end{tabular}
\end{table}

The output from the ADCIRC+SWAN model records every 30 minutes at the locations of the gauges/buoys and errors at each station are computed \loveland{while wind forcing is active to avoid biasing the results to when the contribution from wind waves is not active. This corresponds to computing the error during the 8 day and 18 hour period for Hurricane Ike and the 9 day and 6 hour period for Hurricane Ida as described in Section~\ref{subsec:forcing}.}  \loveland{The WSE from ADCIRC is compared to observations from the NOAA WSE gauges specified in \textbf{Tables~\ref{tab:Elevation_Stations} and \ref{tab:Ida_Elevation_Stations}}, whereas  the wave statistics from SWAN are compared to the observations from the NOAA buoys in \textbf{Table~\ref{tab:Wave_Stations}.}} Two types of error are computed, root mean square error (RMSE) and the percent error of the peak value that are defined as:
\begin{equation}\label{eqn:err}
    \begin{split}
        e_{\rev{\text{RMSE}}} = \sqrt{\frac{\sum_i (u_{i,\text{\rev{sim}}} - u_{i,\rev{\text{meas}}})^2}{N}} \\
        e_{\rev{\text{peak}}} = \frac{ | \max{(u_{sim})} - \max{(u_{\rev{\text{meas}}})} | }{ \max{(u_{\rev{\text{meas}}})}} \times 100.
    \end{split}
\end{equation}
The wave statistics that are recorded at the buoys are significant wave height, peak period, and mean wave direction. Significant wave height is defined as the highest $1/3$ of recorded waves at a point, the peak period is the reciprocal of the frequency where the highest action density is recorded, and mean wave direction is as it sounds. In SWAN, the aforementioned statistical values must be estimated as an \rev{integral quantity of the wave spectrum. There are several references on how to estimate significant wave height from a spectral wave model~\cite{holthuijsen_2007,leblond1981waves,komen_cavaleri_donelan_hasselmann_hasselmann_janssen_1994}.}
The peak period is obtained by finding the period in the spectrum that has the highest action density, and the mean wave direction in degrees is computed by integrating the spectrum weighted by the sine and cosine of the direction of the spectrum~\cite{holthuijsen_2007,leblond1981waves,komen_cavaleri_donelan_hasselmann_hasselmann_janssen_1994}.
\section{Results}\label{sec:Results}
\subsection{Computational Cost}
Total wall clock times rounded to the nearest second are tabulated in ~\textbf{Table~\ref{tab:SWAN_runtimes}}. Each simulation is run in parallel with 1064 of the Intel Xeon Platinum 8280 ("Cascade Lake") processors distributed among 19 computational nodes on the  Frontera supercomputer from the Texas Advanced Computing Center (TACC). \loveland{The choice to use 1064 cores is because it has been found that scaling efficiency diminishes beyond around 2000 nodes per core\rev{, see \cite{tanaka2011scalability}}. For the mesh used for Ike, this means 3150 nodes per core and for Ida, this amounts to 1497 nodes per core \rev{as we observe that new computer architectures with larger cache memory allows near optimal scaling beyond the previously reported 2000 nodes per core and we wish to use the same computational set-up for both hurricanes}.} For the Hurricane Ike scenario ADCIRC without SWAN coupling takes \rev{$1.083$ hours} to run while the fastest ADCIRC+SWAN run is Gen2 with \rev{more than} double the run time at \rev{$2.683$ hours}. Similarly, for Hurricane Ida the ADCIRC without SWAN took  \rev{$0.6434$ hours} hours while the fastest ADCIRC+SWAN run was with Gen2 at \rev{$1.818$ hours}. Thus, we observe in both cases SWAN more than doubles the run time. This overhead is unsurprising because the number of parameters SWAN is solving for at each time step is 505 times larger than ADCIRC. This is because while ADCIRC is solving for 3 quantities (WSE and two velocity components) at each node on the mesh, SWAN is solving for the Action Balance on the whole spectral mesh ($41\times 37$ points) at each node in the finite element mesh. However, SWAN can take much longer time steps ($\sim$10 minutes) compared to ADCIRC ($\sim$1 second). Taking this into consideration, for each SWAN time step of 10 minutes we would expect ADCIRC to have 600 forward solves. Therefore, over the course of a given simulation we would expect SWAN to solve for about $505/600 \approx .85$ times the variables ADCIRC must solve for. In other words, if SWAN were able to solve for each of its unknowns as quick as ADCIRC could, we would expect ADCIRC+SWAN to take 1.85 times the time of ADCIRC by itself. From \textbf{Table~\ref{tab:SWAN_runtimes}} we can see that for any source term package in SWAN it takes considerably longer than 1.85 times just ADCIRC. Thus we can conclude that either SWAN takes longer to solve per unknown than ADCIRC (that would be unsurprising because SWAN solves implicitly while ADCIRC is explicit), there is other associated overhead with running SWAN\loveland{, such as computation of specific source terms,} or a combination of both.

Another interesting part of the results in \textbf{Table~\ref{tab:SWAN_runtimes}} is the large difference in run times due to the choice of source term package. It was observed that running ADCIRC+SWAN with Gen3 source terms took around 1.5 times longer than either Gen1 or Gen2. A plausible explanation for the increased run time is the computational complexity of the Gen3 source terms due to the inclusion of the DIA, the approximation for the four wave nonlinear interactions. \loveland{This could merit further investigation to confirm this intuition but at this time we cannot say for sure this is the sole reason for changes in run time because there are many source terms that change between the \rev{Gen1}/\rev{Gen2} and \rev{Gen3} set ups besides just the DIA.} Additionally, it isn't surprising to see that Gen2 and Gen1 didn't have significantly different run times (only a 5\% difference) because the only difference between the two is the definition of a single parameter in $S_{\rev{\text{in}}}$.
\begin{table}[h!]
\centering
\caption{\label{tab:SWAN_runtimes}  Wall clock times for ADCIRC+SWAN.}
\begin{tabular}{|l|c|c|c|c|}
\hline
{Model Configuration \hspace{6mm}} & { Hurricane Ike Run Time (hr) \hspace{6mm}} & {\rev{Ratio}} & {Hurricane Ida Run Time (hr)} & {Ratio}   \\
\hline
No SWAN & 1.094 & \rev{-} &  0.647 & \rev{-} \\
\hline
Gen1 & 2.837 & \rev{2.593} & 1.881  & \rev{2.907} \\
\hline
Gen2 & 2.688 & \rev{2.457} & 1.818  & \rev{2.809} \\
\hline
Gen3 & 4.086 & \rev{3.735} &  3.206  & \rev{4.955} \\
\hline
\end{tabular}
\end{table}
\subsection{SWAN Output}

\begin{figure}[h!]
  \centering
    \includegraphics[width=\textwidth]{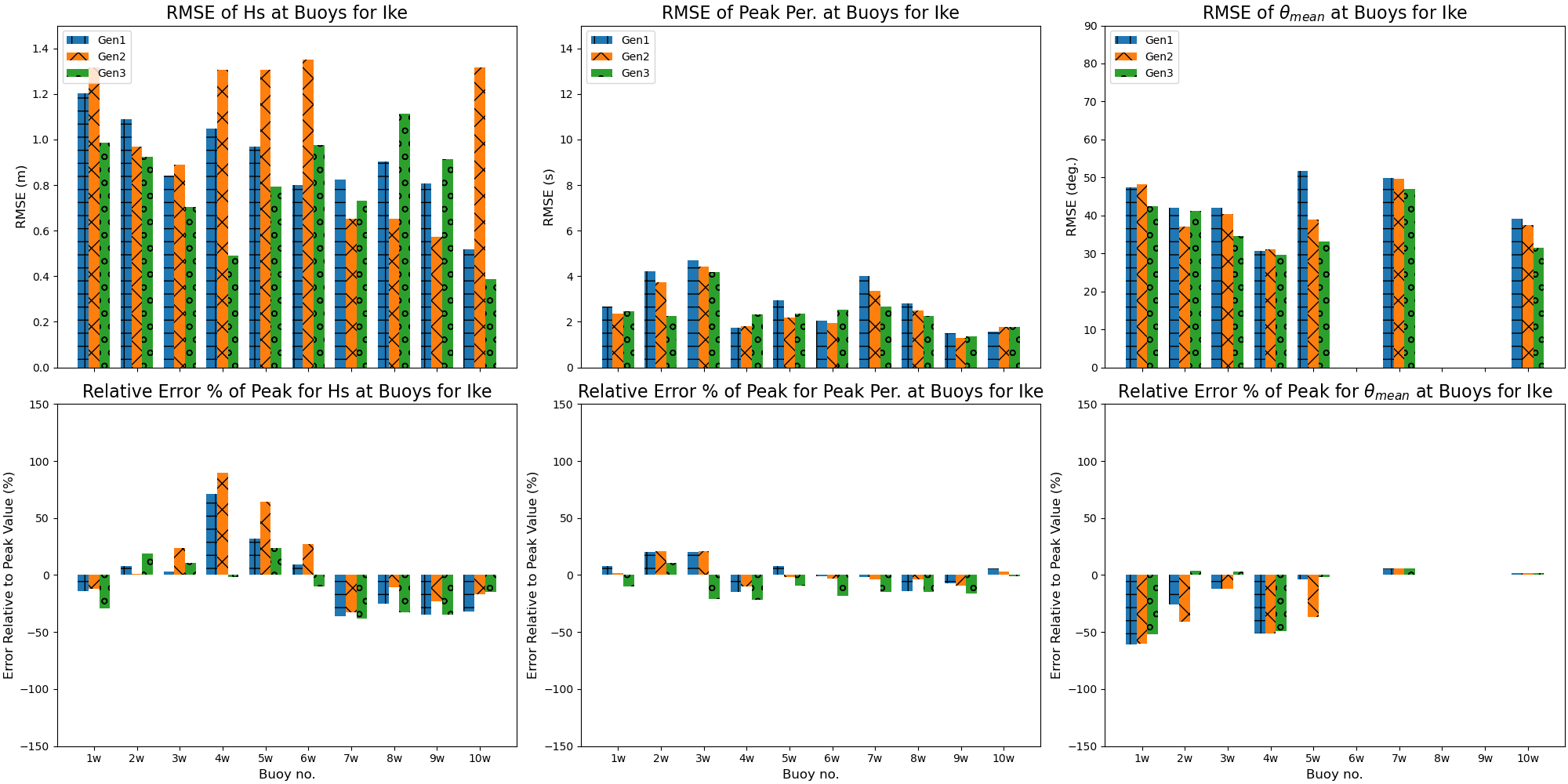}
     \caption{Error statistics relative to NOAA buoy data for Hurricane Ike simulation for significant wave height, peak period, and mean wave directions.}
\label{fig:wave_Ike_err}
\end{figure}

\begin{figure}[h!]
  \centering
    \includegraphics[width=\textwidth]{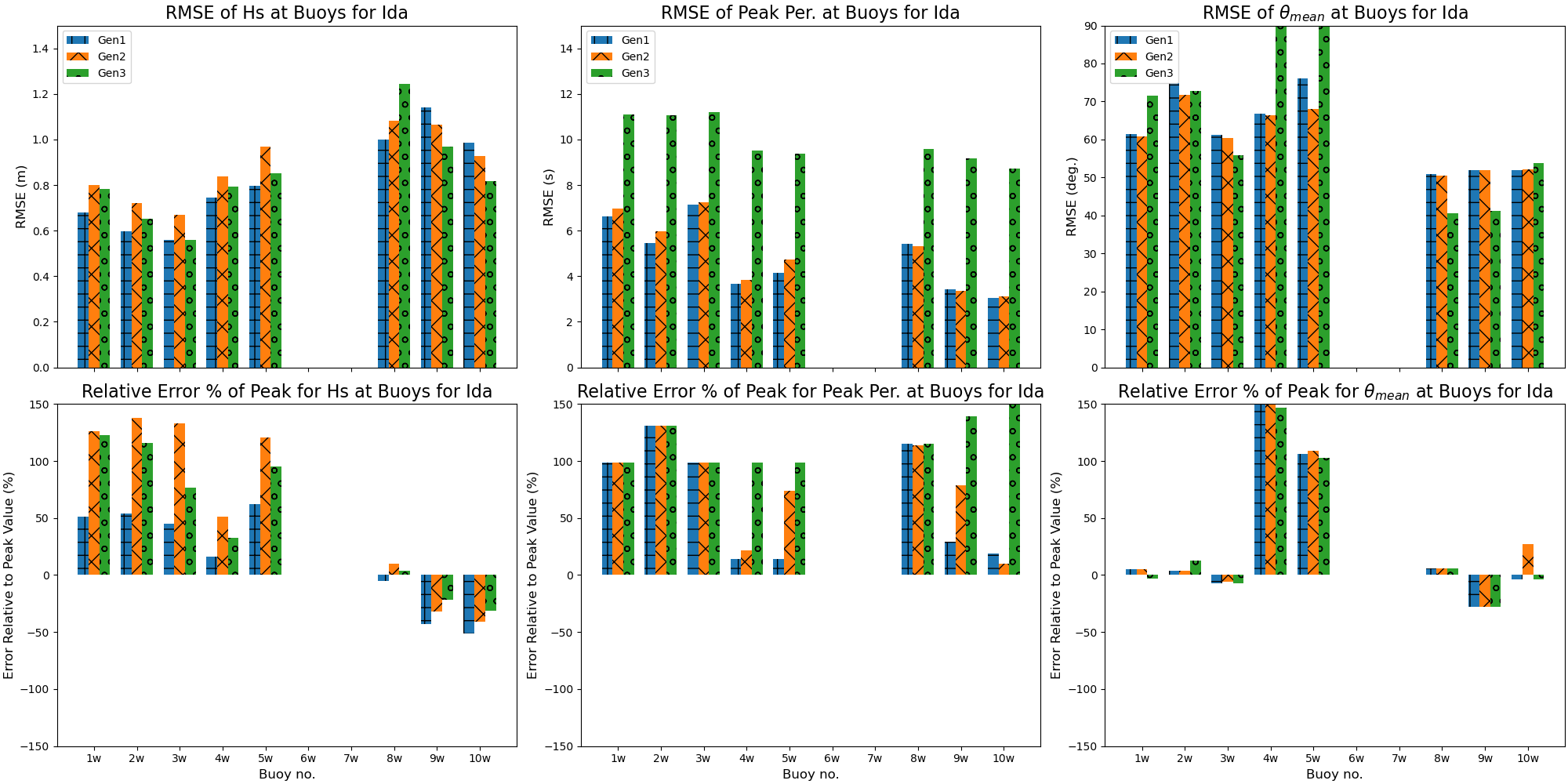}
     \caption{Error statistics relative to NOAA buoy data for Hurricane Ida simulation for significant wave height, peak period, and mean wave directions.}
\label{fig:wave_Ida_err}
\end{figure}

The significant wave height, peak period, and mean wave direction of the ADCIRC+SWAN simulations are used to compute the errors relative to the 10 available NOAA wave buoys from~\textbf{Figure~\ref{fig:wave_gauges}} and plotted as bar graphs for Hurricane Ike in~\textbf{Figure~\ref{fig:wave_Ike_err}} and in~\textbf{Figure~\ref{fig:wave_Ida_err}} for Hurricane Ida.

The average RMSE for significant wave height across all stations for Gen1 is 0.900 m, Gen2 is 1.033 m, and Gen3 is 0.802 meters during Hurricane Ike  and \rev{0}.813 m for Gen1, 0.883 for Gen2, and 0.834 for Gen3 during Hurricane Ida. During the Hurricane Ike  case we do see a 10 \% improvement in RMSE from Gen1 to Gen3 and a roughly 20 \% improvement in error from Gen2 to Gen3 at the buoy locations while during Hurricane Ida we see that Gen1 has the lowest average RMSE but all source term packages are within 0.07 m in average RMSE. The absolute average error percentage relative to the peak value of the buoy was 27 \% for Gen1, 30 \% for Gen2, and Gen3 was 22 \% for Hurricane Ike and for Hurricane Ida was 40 \% for Gen1, 81 \% for Gen2, and 63 \% for Gen3. During Hurricane Ike, much of the difference in the errors between the 3 source term packages was due to buoy number \rev{10w (NBDC 42055)} where both Gen1 and Gen2 nearly doubled the observed peak significant wave height while Gen3 only underpredicted by 2 \%. During Hurricane Ida, it appears that the significant wave height significantly over-predict the observed data except in case of buoys ~\revthree{10w, 8w, and 4w} regardless of source term package.

The average RMSE for peak period across all buoys for Gen1 was 2.822 seconds, Gen2 was 2.541 seconds, and Gen3 was 2.413 seconds for Hurricane Ike  and 4.865 seconds for Gen1, 5.068 seconds for Gen2, and 9.960 seconds for Gen3 during Hurricane Ida. During Hurricane Ike  we do see improved accuracy relative to these 10 buoys as we upgrade in complexity of source terms from Gen1 to Gen2 to Gen3, with a maximal improvement of about 15 \% in error reduction. The absolute average across all buoys of the relative error percentage of the peak is 10 \% for Gen1, 8 \% for Gen2, and  14 \% for Gen3. For Hurricane Ida we find no such improvement with more complex source terms and in fact Gen3 performs significantly worse when compared to buoys.

 The average RMSE for mean wave direction across all buoys is 43.238 degrees for Gen1, 40.392 degrees for Gen2, and 37.076 degrees for Gen3 during Hurricane Ike  and 62.380 degrees for Gen1, 60.229 degrees for Gen2, and 67.73 degrees for Gen3 during Hurricane Ida. During Hurricane Ike , the more complex source terms in Gen3 do improve the RMSE over the buoys and in this case by roughly 14 \% relative to Gen1. The absolute average across all buoys of the relative error percentage to the peak buoy value is 23 \% for Gen1, 30 \% for Gen2, and 17 \% for Gen3. However, no such improvements are observed during Hurricane Ida. 

\begin{landscape}
    \begin{figure}[h!]
    \centering
    \includegraphics[width=8.5in]{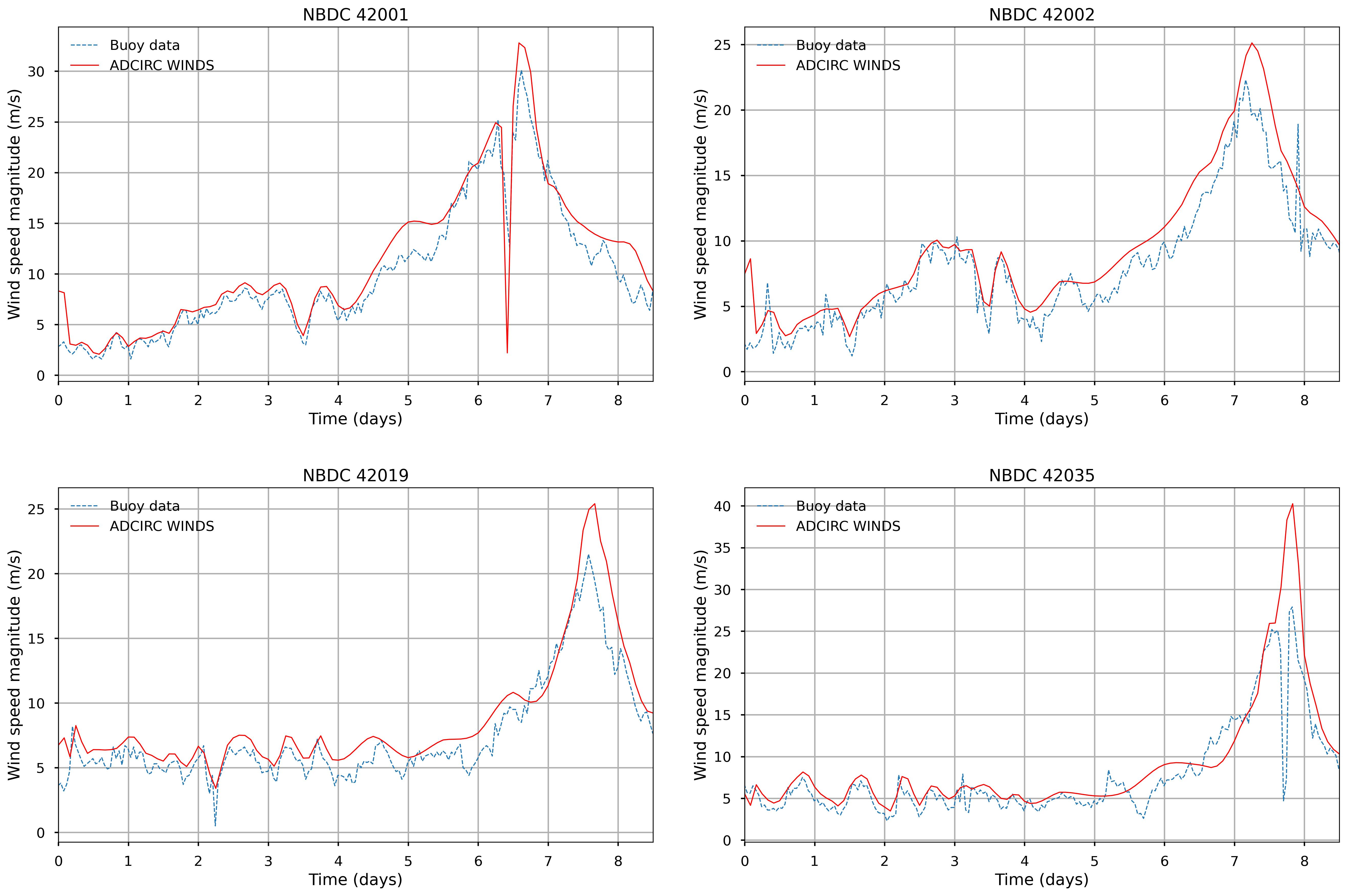}
    \caption{Wind velocity magnitude of NOAA buoys near track along with ADCIRC+SWAN wind forcing during Hurricane Ike. \textbf{Top left:} Buoy 6w. \textbf{Top right:} Buoy 5w. \textbf{Bottom left:} Buoy 2w. \textbf{Bottom right:} Buoy 3w.}
    \label{fig:Ike_winds_1}
\end{figure}
\end{landscape}

\begin{landscape}
    \begin{figure}[h!]
    \centering
    \includegraphics[width=8.5in]{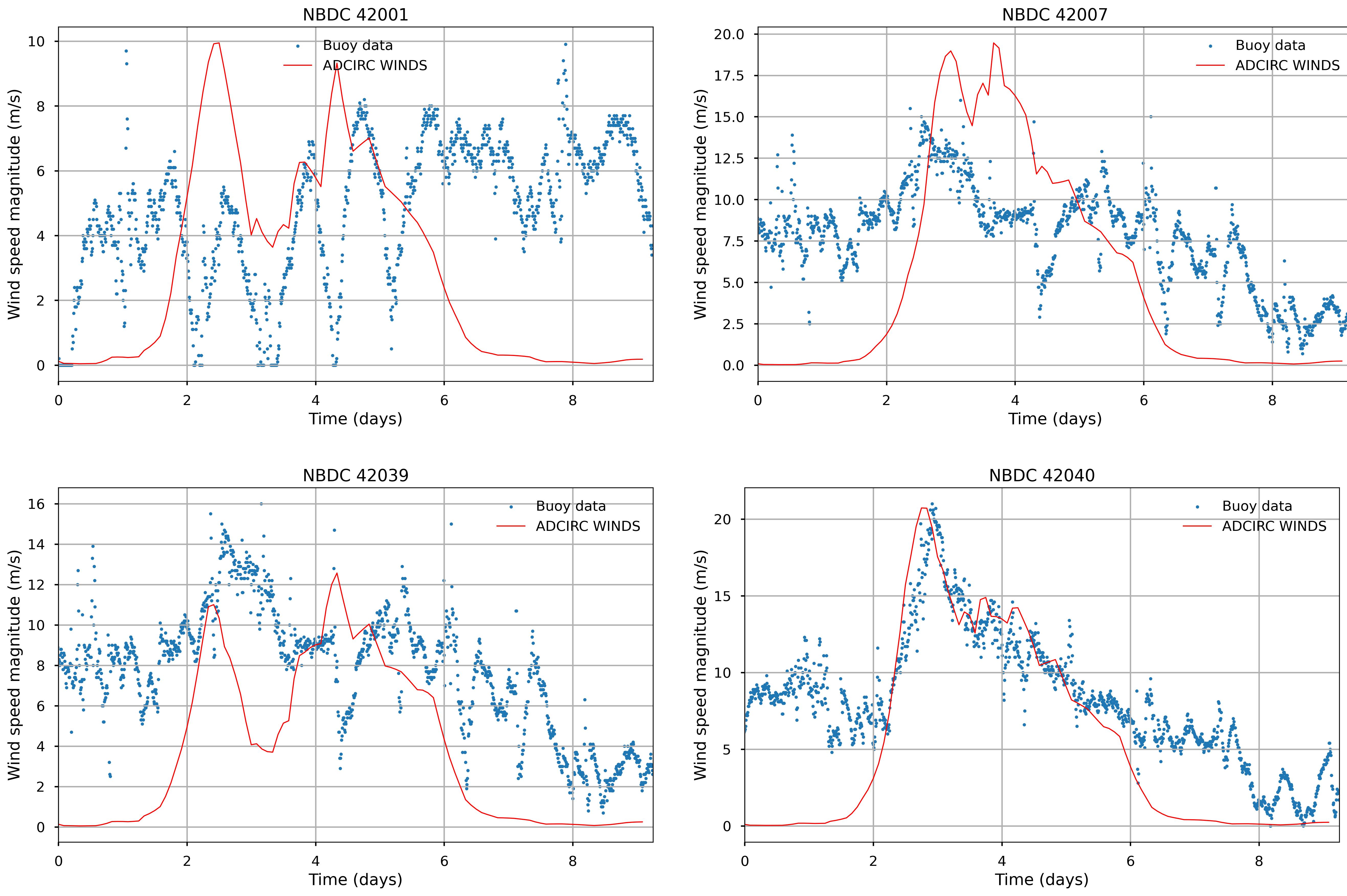}
    \caption{Wind velocity magnitude of NOAA buoys near track along with ADCIRC+SWAN wind forcing during Hurricane Ida. \textbf{Top left:} Buoy 6w. \textbf{Top right:} Buoy 7w. \textbf{Bottom left:} Buoy 9w. \textbf{Bottom right:} Buoy 8w.}
    \label{fig:Ida_winds_2}
\end{figure}
\end{landscape}

The larger errors in some of the SWAN output with respect to buoys are most likely related to the inaccuracy of the wind fields, particularly during Hurricane Ida. Because the best track winds were used for Hurricane Ida, this results in a  simplified wind field that doesn't capture the irregular oscillations that are observed by the buoys as shown in~\textbf{Figure~\ref{fig:Ida_winds_2}}. The average RMSE in wind magnitude across all buoys is roughly 5 m/s during Hurricane Ida while it is only around 2 m/s during Hurricane Ike. A comparison between wind velocity magnitude of the ADCIRC simulation and the buoy data during Hurricane Ike is shown in~\textbf{Figure~\ref{fig:Ike_winds_1}}.

\begin{figure}[h!]
  \centering
    \includegraphics[width=\textwidth]{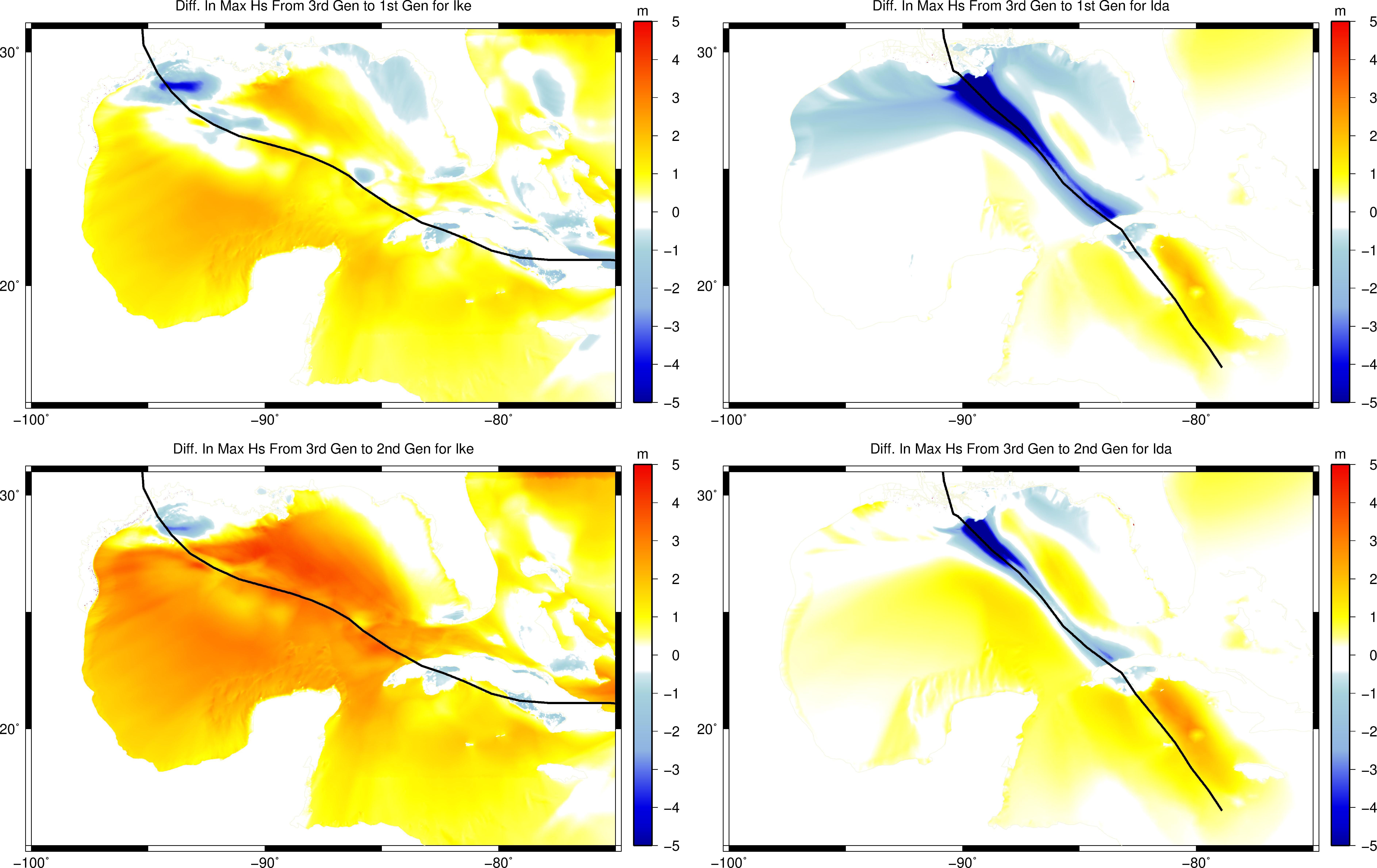}
     \caption{Differences between maximum significant wave heights due to choices of source term package. \textbf{Top left:} Max $H_s$ of \rev{Gen1} minus max $H_s$ of \rev{Gen3} for Ike. \textbf{Top right:} Max $H_s$ of \rev{Gen1} minus max $H_s$ of \rev{Gen3} for Ida. \textbf{Bottom left:} Max $H_s$ of \rev{Gen2} minus max $H_s$ of \rev{Gen3} for Ike. \textbf{Bottom right:} Max $H_s$ of \rev{Gen2} minus max $H_s$ of \rev{Gen3} for Ida.}
\label{fig:HS_compare}
\end{figure}

\loveland{We can also look at global sensitivity of the wave statistics outputs in ADCIRC+SWAN related to the choice of source terms. In ~\textbf{Figure~\ref{fig:HS_compare}} we can see the differences in maximum significant wave height due to changes in the source term configurations for both Ike and Ida simulations. We can see that during Ike, the significant wave height globally seems to be lower in the deep ocean when compared to \rev{Gen1} or \rev{Gen2} but near the coast and near the hurricane track \rev{Gen3} has maximum significant wave heights that are a meter or more higher than the reduced order scenarios (\textbf{Figure~\ref{fig:HS_compare}}, top left and bottom left). During Hurricane Ida, it is interesting to see that near the track the max significant wave height is multiple meters higher for the \rev{Gen3} than either the \rev{Gen1} or \rev{Gen2} scenarios (\textbf{Figure~\ref{fig:HS_compare}}, top right and bottom right).}

%
%

\subsection{ADCIRC Output}

\begin{figure}[h!]
  \centering
    \includegraphics[width=\textwidth]{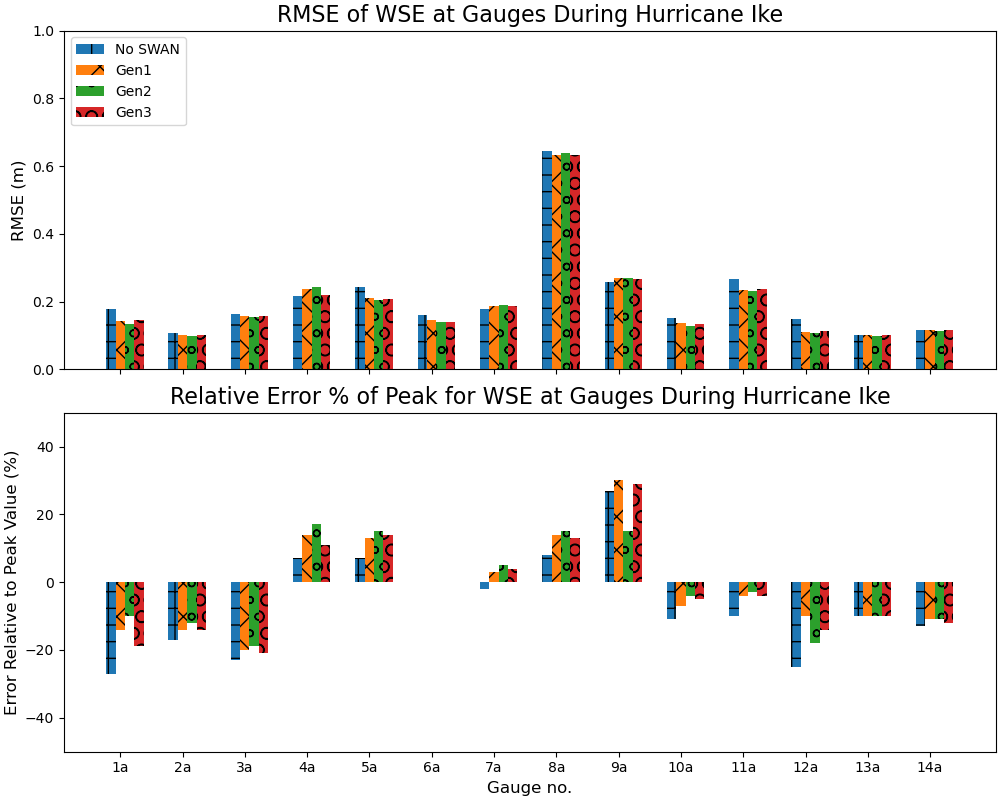}
     \caption{Error statistics relative to NOAA gauge data for Hurricane Ike simulation.}
\label{fig:wse_Ike_err}
\end{figure}

\begin{figure}[h!]
  \centering
    \includegraphics[width=\textwidth]{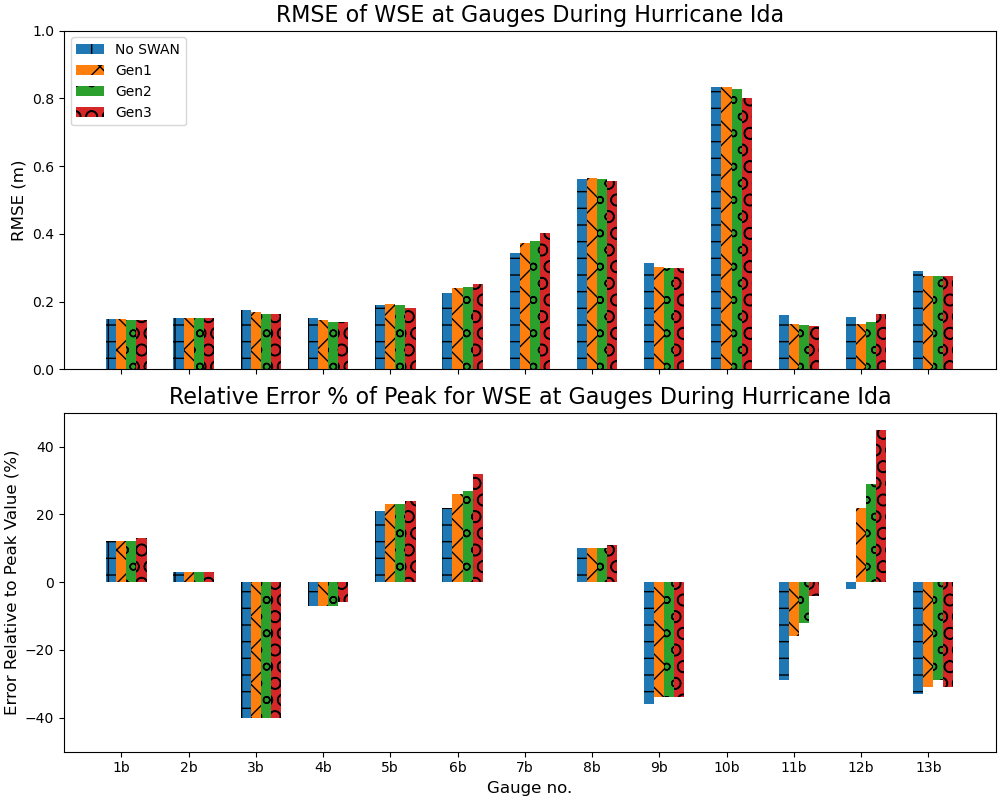}
     \caption{Error statistics relative to NOAA gauge data for Hurricane Ida simulation.}
\label{fig:wse_Ida_err}
\end{figure}

\loveland{During the Hurricane Ike simulation, the data from the 14 NOAA gauges along the Texas coast (see ~\textbf{Figure~\ref{fig:ele_gauge_all}}), and the output WSE from ADCIRC+SWAN are used to compute the errors, $e_{\rev{\text{RMSE}}}$ and $e_{\rev{\text{peak}}}$ as in~\eqref{eqn:err}. The error statistics are plotted in bar graphs to compare the 4 configurations (No SWAN, \rev{Gen1}, \rev{Gen2}, \rev{Gen3}) in ~\textbf{Figure~\ref{fig:wse_Ike_err}}.
Similarly, during the Hurricane Ida simulation, the 13 NOAA gauges along the Louisiana coast (see~\textbf{Figure~\ref{fig:Ida_gauge_locs}}), and the output WSE from ADCIRC+SWAN are used to compute the errors \rev{that} are plotted for all of the 4 configurations in ~\textbf{Figure~\ref{fig:wse_Ida_err}}. During Hurricane Ike, the average RMSE across all gauges when excluding waves altogether is 0.210 m while for Gen1 it is \rev{0}.190 m, Gen2 is \rev{0}.196 m, and Gen3 is \rev{0}.197 m. During Hurricane Ida, the average  RMSE across all gauges when excluding waves altogether is 0.285 meters while for Gen1 it is \rev{0}.282 m, Gen2 is \rev{0}.281 m, and Gen3 is \rev{0}.282 m.}

Including waves resulted in less than a 5 \% reduction in RMSE during Hurricane Ike and less than 1.5 \% during Hurricane Ida at the gauge locations. The choice of source terms really did not impact RMSE with only a maximum difference of .007 m between Gen1 and Gen3 during Hurricane Ike and a maximum difference of \rev{0}.001 m between Gen2 and Gen3. The absolute average relative error percent to the peak value without accounting for waves is about 13 \% while Gen1 is 12 \%, Gen2 is 10 \% , and Gen3 is 11 \% during Hurricane Ike. For Hurricane Ida, the same quantity is 20 \% without waves while Gen1 is 20 \%, Gen2 is 21 \%, and Gen3 is 22 \%. Qualitatively examining the outputs in the graphs shown in~\textbf{Figure~\ref{fig:wse_Ike_err}} and ~\textbf{Figure~\ref{fig:wse_Ida_err}} does not show any drastic improvements in accuracy between one source term package over the others across all the gauge locations and during Hurricane Ida it doesn't appear that including waves reduces the error compared to the NOAA WSE gauge data overall. This is not to say including waves does not improve the overall quality of the simulation.

\begin{figure}[h!]
  \centering
    \includegraphics[width=\textwidth]{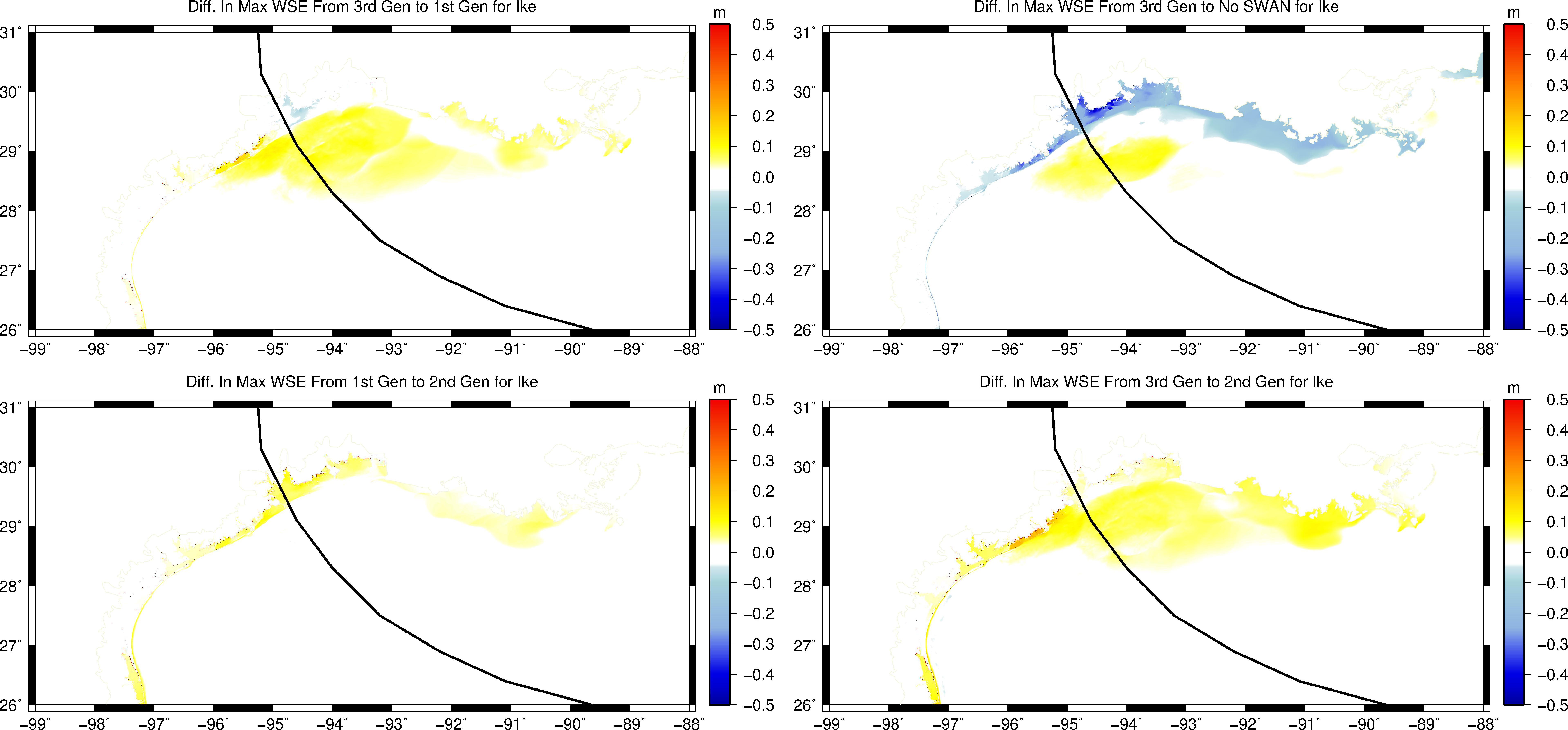}
     \caption{Differences between maximum water surface elevation due to choices of source term package. \textbf{Top left:} Max WSE of \rev{Gen1} minus max WSE of \rev{Gen3}. \textbf{Top right:} Max WSE of No SWAN minus max WSE of \rev{Gen3}. \textbf{Bottom left:} Max WSE of \rev{Gen12} minus max WSE of \rev{Gen1}. \textbf{Bottom right:} Max WSE of \rev{Gen2} minus max WSE of \rev{Gen3}. }
\label{fig:Ike_WSE_compare}
\end{figure}

\begin{figure}[h!]
  \centering
    \includegraphics[width=\textwidth]{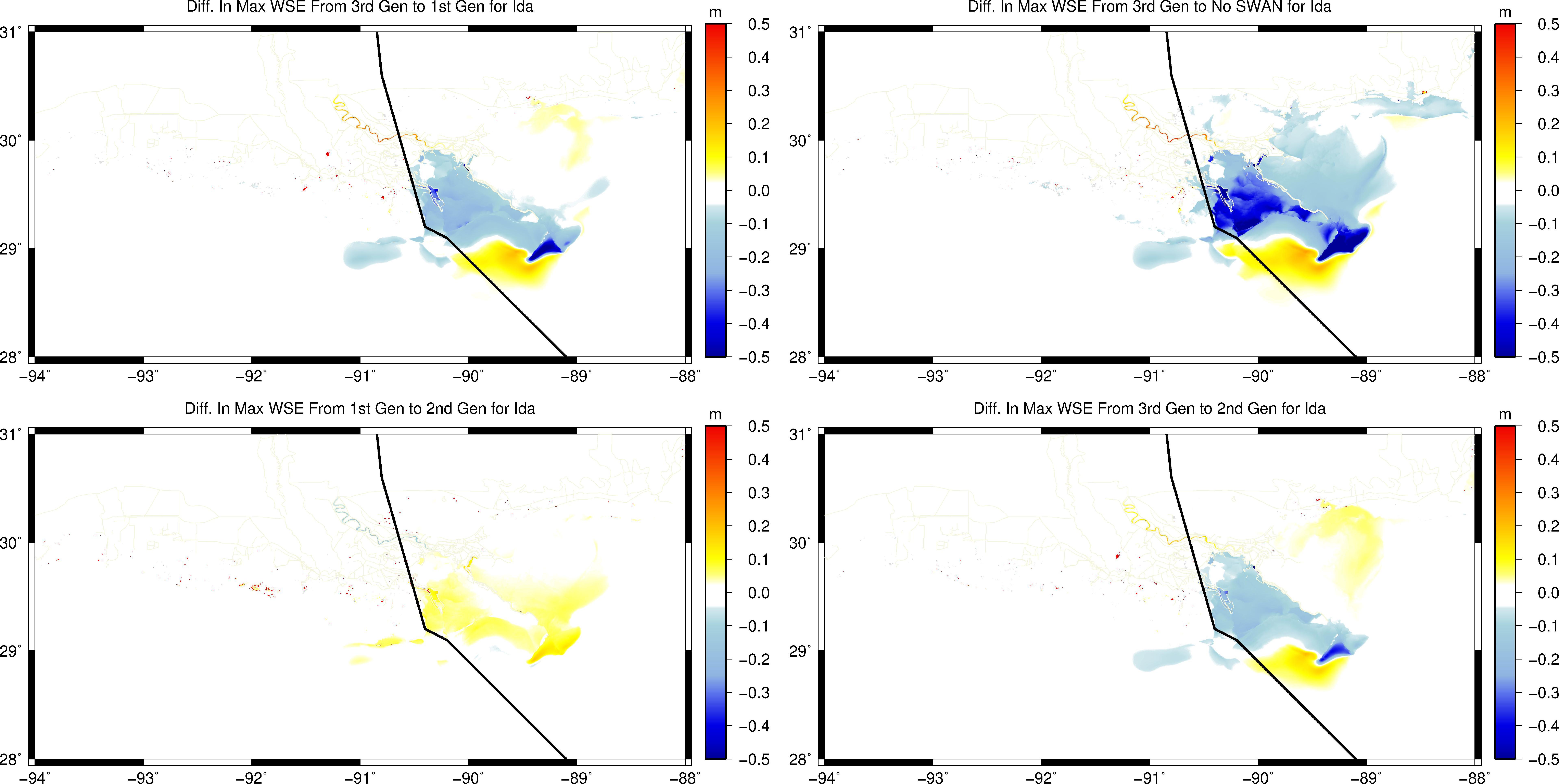}
     \caption{Differences between maximum water surface elevation due to choices of source term package. \textbf{Top left:} Max WSE of \rev{Gen1} minus max WSE of \rev{Gen3}. \textbf{Top right:} Max WSE of No SWAN minus max WSE of \rev{Gen3}. \textbf{Bottom left:} Max WSE of \rev{Gen2} minus max WSE of \rev{Gen1}. \textbf{Bottom right:} Max WSE of \rev{Gen2} minus max WSE of \rev{Gen3}. }
\label{fig:Ida_WSE_compare}
\end{figure}

We can see that outside of analyzing the accuracy of time series WSE data at the gauge locations, there is a large sensitivity to global ADCIRC+SWAN WSE levels due to source term configuration within SWAN. In~\textbf{Figure~\ref{fig:Ike_WSE_compare}} we have plotted the differences in the maximum recorded WSE from ADCIRC+SWAN of the various source term configurations during Hurricane Ike, and we have the same quantities plotted for Hurricane Ida in~\textbf{Figure~\ref{fig:Ida_WSE_compare}}. For Hurricane Ike, it can be seen that excluding waves altogether leads to a reduction of max WSE of nearly 0.5 m near the coastline (\textbf{Figure~\ref{fig:Ike_WSE_compare}}, top right) and for Ida we see similar results with the exclusion of waves reducing max WSE near the coastline by nearly 0.5 m (\textbf{Figure~\ref{fig:Ida_WSE_compare}}, top right).

For Ike, it appears that the max WSE of the \rev{Gen3} simulation is nearly 0.1 m less on average in the vicinity of the coast when compared to both \rev{Gen1} and \rev{Gen3} simulations (\textbf{Figure~\ref{fig:Ike_WSE_compare}}, top left and bottom right respectively). For Ida, it appears that the max WSE of the \rev{Gen3} simulation varies spatially and on average the max WSE seems to be higher than both \rev{Gen1} and \rev{Gen2} simulations, though there are significant regions near the track of Ida where \rev{Gen1} and \rev{Gen2} have higher max WSE compared to \rev{Gen3} (\textbf{Figure~\ref{fig:Ike_WSE_compare}}, top left and bottom right respectively). We can also see there are differences in max WSE between the reduced order source terms in both Ida and Ike, though much smaller in magnitude than between any of the other simulations (\textbf{Figure~\ref{fig:Ike_WSE_compare}} and \textbf{Figure~\ref{fig:Ida_WSE_compare}}, bottom left).

\rev{We also note that for both storms, we see the largest errors with respect to gauge data as well as the largest differences between each simulation occur closest to the track of the storm. This is unsurprising because the largest magnitudes in winds occur in the vicinity of the track. The higher winds will increase WSE and wave action in the area resulting in potential errors due to inaccuracies in factors such as the wind field itself, or physics parameterizations in ADCIRC or SWAN to be larger.}

\section{Conclusions}\label{sec:con}
From this study, it was observed that varying the source term packages in SWAN as part of a validated ADCIRC+SWAN model significantly impacted run times as well as results in wave outputs. Upgrading from Gen1 or Gen2 source terms to the ST6 Gen3 source terms resulted in about a 40 \% increase in run time. The effects of the choice of source terms (Gen1, Gen2 or Gen3) on average WSE at NOAA gauges only changed RMSE relative to the gauges by about \rev{0}.007 m. However, more significant differences in wave statistics were observed at the NOAA buoy locations based on source term choice. These differences greatly depended on the scenario, as during Hurricane Ike  Gen3 source terms showed roughly a 20 \% improvement in the accuracy of significant wave height, and a 15 \% improvement in accuracy of peak period, and a 14 \% improvement in accuracy mean wave direction relative to Gen1 or Gen2 source terms while during Hurricane Ida no such improvements in accuracy were observed.

This study shows the possible trade-off between accuracy and run time for the choice of source term complexity in SWAN in the coupled ADCIRC+SWAN setting. The small differences in accuracy compared to observed water levels of the source term configurations may warrant more investigation into the use of reduced order source term packages such as Gen1 or Gen2. For instance, it may be worth the savings in computation if only water surface elevations are of primary interest as opposed to wave statistics. \loveland{However, large sensitivities to the source term choice in global output of significant wave height near hurricane tracks were observed.} \loveland{A key observation from this study is that if the meteorological forcing isn't \rev{sufficiently} accurate, which is common in forecasting scenarios, then the additional computational cost associated with the  detailed Gen3 source terms may not improve accuracy of the model.} \rev{If there is access to highly accurate wind fields, as was the case for the Ike simulation in this study, then Gen3 source terms should yield improved accuracy in WSE and wave statistics when compared to Gen1 or Gen2.}  \loveland{Additionally, the computational savings shown when using the reduced order source terms could allow for potentially higher spectral resolutions to be used.} 

Some limitations of this study is that this was for a single geographic region, only two storm were run, and the source terms were not specifically tuned for this region so these results may vary significantly for a different domain or accuracy of the wave model could be improved if the source terms were tuned specifically for this region.

\section*{Acknowledgements} 
This work has been supported by the United States Department of Homeland Security Coastal Resilience Center research project "Accurate and Fast Wave Modeling and Coupling with ADCIRC".
The authors also would like to gratefully acknowledge the use of the"DMS23001" and "DMS21031" allocations on the Frontera supercomputer at the Texas Advanced Computing Center at the University of Texas at Austin.

\appendix
\section{Input Files for ADCIRC+SWAN}\label{ap:Source}
\noindent \revthree{The following text shows the exact fort.26 file that was used for the Gen1 simulation for Hurricane Ida:}
\begin{lstlisting}[
    basicstyle= \small
]
$*************************HEADING************************
$
PROJ 'IdaWW' 'GWW'
$
$********************MODEL INPUT*************************
$
SET LEVEL 0.0
SET DEPMIN 0.1
MODE NONSTATIONARY
$
COORDINATES SPHERICAL CCM
$
CGRID UNSTRUCTURED CIRCLE MDC=36 FLOW=0.031384 MSC=40
READ UNSTRUCTURED
$
INIT ZERO
$
INPGRID WLEV UNSTRUCTURED EXCEPTION 0.1  NONSTAT 20210826.120000 600 SEC 20210904.180000
READINP ADCWL
$
INPGRID  CUR UNSTRUCTURED EXCEPTION 0.   NONSTAT 20210826.120000 600 SEC 20210904.180000
READINP ADCCUR
$
INPGRID WIND UNSTRUCTURED EXCEPTION 0.   NONSTAT 20210826.120000 600 SEC 20210904.180000
READINP ADCWIND
$
INPGRID FRIC UNSTRUCTURED EXCEPTION 0.05 NONSTAT 20210826.120000 600 SEC 20210904.180000
READINP ADCFRIC
$
GEN1
PROP BSBT
NUM STOPC DABS=0.005 DREL=0.01 CURVAT=0.005 NPNTS=95 NONSTAT MXITNS=20
$
$*************************************************************
$
QUANTITY HS DIR TMM10 TPS FMIN=0.031384 FMAX=1.420416
$
TEST 1,0
COMPUTE 20210826.120000 600 SEC 20210904.180000
STOP
$
\end{lstlisting}
\revthree{Next we have the exact fort.26 used in the Gen2 set up. This file is identical to the first except for the line specifying the source package changes from Gen1 to Gen2:}
\begin{lstlisting}[
    basicstyle= \small
]
$*************************HEADING************************
$
PROJ 'IdaWW' 'GWW'
$
$********************MODEL INPUT*************************
$
SET LEVEL 0.0
SET DEPMIN 0.1
MODE NONSTATIONARY
$
COORDINATES SPHERICAL CCM
$
CGRID UNSTRUCTURED CIRCLE MDC=36 FLOW=0.031384 MSC=40
READ UNSTRUCTURED
$
INIT ZERO
$
INPGRID WLEV UNSTRUCTURED EXCEPTION 0.1  NONSTAT 20210826.120000 600 SEC 20210904.180000
READINP ADCWL
$
INPGRID  CUR UNSTRUCTURED EXCEPTION 0.   NONSTAT 20210826.120000 600 SEC 20210904.180000
READINP ADCCUR
$
INPGRID WIND UNSTRUCTURED EXCEPTION 0.   NONSTAT 20210826.120000 600 SEC 20210904.180000
READINP ADCWIND
$
INPGRID FRIC UNSTRUCTURED EXCEPTION 0.05 NONSTAT 20210826.120000 600 SEC 20210904.180000
READINP ADCFRIC
$
GEN2
PROP BSBT
NUM STOPC DABS=0.005 DREL=0.01 CURVAT=0.005 NPNTS=95 NONSTAT MXITNS=20
$
$*************************************************************
$
QUANTITY HS DIR TMM10 TPS FMIN=0.031384 FMAX=1.420416
$
TEST 1,0
COMPUTE 20210826.120000 600 SEC 20210904.180000
STOP
$
\end{lstlisting}
\revthree{Lastly, we have the fort.26 that was used for Gen3 source terms during Hurricane Ida. We note that the fort.26 files for the Hurricane Ike runs were similar except for the date ranges.}
\begin{lstlisting}[basicstyle= \small]
$*************************HEADING************************
$
PROJ 'IdaWW' 'GWW'
$
$********************MODEL INPUT*************************
$
SET LEVEL 0.0
SET DEPMIN 0.1
MODE NONSTATIONARY
$
COORDINATES SPHERICAL CCM
$
CGRID UNSTRUCTURED CIRCLE MDC=36 FLOW=0.031384 MSC=40
READ UNSTRUCTURED
$
INIT ZERO
$
INPGRID WLEV UNSTRUCTURED EXCEPTION 0.1  NONSTAT 20210826.120000 600 SEC 20210904.180000
READINP ADCWL
$
INPGRID  CUR UNSTRUCTURED EXCEPTION 0.   NONSTAT 20210826.120000 600 SEC 20210904.180000
READINP ADCCUR
$
INPGRID WIND UNSTRUCTURED EXCEPTION 0.   NONSTAT 20210826.120000 600 SEC 20210904.180000
READINP ADCWIND
$
INPGRID FRIC UNSTRUCTURED EXCEPTION 0.05 NONSTAT 20210826.120000 600 SEC 20210904.180000
READINP ADCFRIC
$
GEN3 ST6 4.70E-7 6.6E-6 4 4 UP HWANG VECTAU U10PROXY 28 AGROW
SSWELL ARDHUIN 1.2
WCAP KOMEN 2.36E-5 3.02E-3 2.0 1.0 1.0
BREAKING
FRICTION MADSEN KN=0.05
PROP BSBT
NUM STOPC DABS=0.005 DREL=0.01 CURVAT=0.005 NPNTS=95 NONSTAT MXITNS=20
$
$*************************************************************
$
QUANTITY HS DIR TMM10 TPS FMIN=0.031384 FMAX=1.420416
$
TEST 1,0
COMPUTE 20210826.120000 600 SEC 20210904.180000
STOP
$
\end{lstlisting}

\printcredits

\bibliographystyle{cas-model2-names}

\bibliography{bibliography}



\end{document}